\documentstyle[proceedings]{crckapb}
\input{epsf.sty}
\def\figdir{.}

\def\note #1]{{\bf #1]}}
\def\etal{{\it et al.}}
\def\eg{{\it e.g.}}
\def\cf{{\it cf.}}
\def\ie{{\it i.e.}}

\def\muHz{\,\mu{\rm Hz}}
\def\K{\,{\rm K}}
\def\rt{r_{\rm t}}
\begin{opening}
\title{CONSTRAINTS ON STELLAR INTERIOR PHYSICS FROM HELIOSEISMOLOGY}
\author{J. Christensen-Dalsgaard}
\institute{Teoretisk Astrofysik Center, Danmarks Grundforskningsfond, 
and
Institut for Fysik og Astronomi, Aarhus Universitet, \\
DK-8000 Aarhus C, Denmark}
\end{opening}
\runningtitle{CONSTRAINTS FROM HELIOSEISMOLOGY}
\begin{document}

\section{Introduction}

Traditional observations of the properties of stars generally provide
tests of only the gross aspects of stellar structure and evolution.
%Important examples are the distribution of
%stars in the HR diagram of open clusters \note [ref??],
%or the properties of stars in well-observed binary systems
%(e.g. \note [Andersen]).
%Such data generally show a reasonable agreement with the results
%of traditional stellar evolution calculations, although evidence
%has been found for effects which might be considered nonstandard
%processes, such as overshoot from convective cores (e.g. \note [ref]).
%
The limitation lies in the amount and precision of the available data 
%in traditional observations 
of relevance to the structure of the
stellar interior, {\ie}, the determination of stellar effective
temperatures, surface composition, luminosities and, in a few cases, masses.
Additional constraints on the observed stars,
such as the common age and composition normally assumed 
for stars in clusters or multiple systems,
clearly increase the information. % that can be obtained.
%as does the 
%dynamical behaviour of binary systems, as observed in absidal
%motion ({\eg} \note [ref]).
However, detailed information on the physics and processes of 
stellar interiors requires more extensive data, with a dependence
on stellar structure sufficiently simple to allow unambiguous 
interpretation.
Such data are offered by observations of stellar pulsation frequencies:
they can be observed with great accuracy and their dependence
of stellar structure is generally well understood.
%Hence they provide stringent constraints on stellar models.
%Even observations of just two pulsation periods in some
%Cepheid and $\delta$ Scuti stars provided the impetus for
%posed severe problems for 
%stellar evolution theory ({\eg} \note [refs]),
%which were only satisfactorily resolved with 
%a major revision of stellar opacities ({\eg} \note [refs]).
%However, it is 
In particular, the richness and precision of the observed
frequencies of solar oscillation are now offering a detailed
view on the interior properties of a star.

Reviews on solar oscillations were
provided by, {\eg}, Gough \& Toomre (1991).
The modes are characterized by a degree $l$ measuring,
approximately, the number of wavelengths in the stellar circumference.
%Modes with $l = 0$ are spherically symmetric and correspond to
%the radial oscillations normally assumed to be responsible for ``classical''
%stellar pulsations, such as those of Cepheids.
For each $l$, there is a set of possible modes of oscillation,
characterized by the radial order $n$, and with angular frequencies
$\omega_{nl}$.
In a spherically symmetric model the frequencies are
independent of the azimuthal order $m$.
This degeneracy is lifted by
rotation or other departures from spherical symmetry.

In the Sun, modes are observed at each $l$ between $l = 0$ and
at least 2000, with cyclic frequencies
$\nu = \omega/2 \pi$ between about $1000$ and $5000 \muHz$.
The relative standard deviations 
are less than $5\times 10^{-6}$ in many cases, making these 
frequencies by far the most accurately known properties of the Sun.
The observed modes are essentially standing
acoustic waves, propagating between a point just below the
photosphere and an inner turning point at a distance $\rt$
from the centre such that $c(\rt)/\rt = \omega / \sqrt{l(l+1)}$,
$c$ being the adiabatic sound speed.
Over the observed range of modes
$\rt$ moves from very near the centre of the Sun to just below
the solar surface.
This variation of the region to which 
the mode frequencies are sensitive permits inverse analyses
%discriminating between the influence of different parts of the Sun 
to determine localized properties of the solar interior.

The analysis of the observed frequencies is generally 
aimed at determining differences between the Sun and reference solar models
and hence inferring the errors in the assumed physics or other
properties of the models.
The quality of the helioseismic data has inspired 
considerable efforts to improve the solar model computations,
by including as far as possible known processes and by using 
the most precise description of the physics available.
In particular, diffusion and gravitational settling, 
which have often been neglected in the past,
have a substantial effect on the models at this level of precision 
and hence must be included.

Here I shall use as reference the so-called Model S of
Christensen-Dalsgaard {\etal} (1996), which includes 
diffusion and settling of helium and heavy elements;
OPAL equation of state (Rogers, Swenson \& Iglesias 1996) and
opacity (Iglesias, Rogers \& Wilson 1992) were used,
as well as nuclear reaction rates
largely from Bahcall \& Pinsonneault (1995).
This, as well as other models discussed here,
were calibrated to present solar luminosity and radius,
as well as to the observed surface ratio $Z_{\rm s}/X_{\rm s} = 0.0245$
between the heavy-element and hydrogen abundances,
by adjusting the initial composition and the mixing-length parameter.

\section{Results of helioseismic inversion}

The differences between the observed and model frequencies are small,
of order 0.3 \% or less.
This motivates analysis in terms of linearized relations 
between the frequency differences and differences in suitable sets of
model variables.
Here I use $(c^2, \rho)$, $\rho$ being density.
It is possible to form linear combinations of the frequency differences,
in such a way as to obtain a measure of the sound-speed difference
$\delta c/c$, localized to a small region of the Sun, while suppressing the
influence of the difference in $\rho$, of uncertainties
in the modelling of the near-surface region of the Sun
and of observational errors (for details, see for example
Basu {\etal} 1996).
%Thus we can determine the error in the model sound speed as
%a function of position in the Sun.
Differences between the Sun and Model S obtained in this manner
are shown in Fig.~1,
based on frequencies resulting from a combination of observations from
the BiSON network (Chaplin {\etal} 1996) and
the LOWL instrument (Tomczyk {\etal} 1995).
%for details, see Basu {\etal} (1997).
As indicated by the horizontal bars, the points provide averages
of $\delta c^2/c^2$ over limited regions in the Sun, from the centre 
to near the surface.
Furthermore, the estimated random errors in the result, based on the
quoted errors in the observed frequencies, are minute,
less than $2\times 10^{-4}$ in most of the solar interior.
%Test have shown that the resulting sound speed depends little 
%on the choice of reference model.
Thus it is in fact possible to measure a property of the solar interior
as a function of position, with great precision.
%Needless to say, this provides stringent constraints on the solar models
%and hence on the underlying physics and assumptions.

\begin{figure}[ht]
\epsfxsize=9cm\hspace*{1.8cm}\epsfbox{\figdir/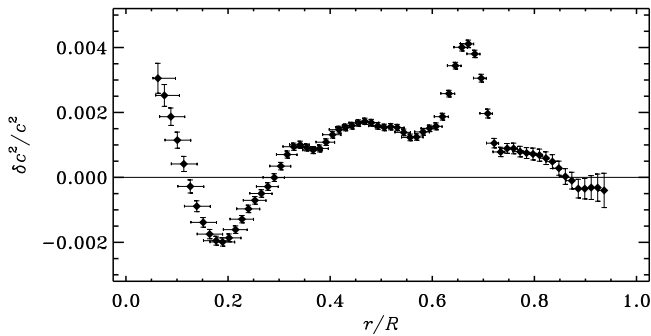}
\vskip -0.3cm
\caption{Inferred difference in squared sound speed,
in the sense (Sun) -- (model).
The horizontal error bars mark the first and third
quartile points of the averaging kernels,
whereas the vertical error bars show 1-$\sigma$ errors,
as progated from the errors in the observed frequencies.
From Basu {\etal} (1997).}
\end{figure}

It is evident, also, that at a superficial level the agreement between the 
Sun and the model is excellent: 
%by incorporating the best available physics in the calculation 
we have been able to predict the
solar sound speed with a precision of better than 0.2 \%.
This has required improvements in the modelling inspired
by the high accuracy of the observations;
however, the calculation involves no adjustment of parameters
to fit the model to the data.
On the other hand, the difference between
the Sun and the model is highly significant, given the very small
error in the inferred difference.
Thus, in that sense, the model is hardly satisfactory.

%Helioseismic inversion also yields the
%density difference $\delta \rho/\rho$ between the Sun and the model;
%this shows that the model reproduces the solar density profile
%to better than approximately 1 \%, the biggest difference being
%in the convection zone ({\eg} Gough {\etal} 1996;
%Basu {\etal} 1997).

The oscillations depend essentially
only on the dynamical properties of the Sun, {\eg},
pressure, density and sound speed.
%relating the pressure and density perturbations.
%On the other hand, there is no direct dependence on temperature $T$.
Since, approximately, $c^2 \propto T/\mu$,
where $T$ is temperature and $\mu$ the mean molecular weight,
helioseismology constrains $T/\mu$ but not $T$ and $\mu$ separately.
This has ramifications for the possibility of helioseismic
constraints on the solar neutrino production
({\eg} Antia \& Chitre 1995; Christensen-Dalsgaard 1997).

The dependence of the oscillation frequencies
on azimuthal order $m$ carries information about the solar internal
angular velocity.
% $\Omega(r, \theta)$, $\theta$ being co-latitude.
%It is possible to invert this dependence to infer the variation
%of $\Omega$ with $r$ and $\theta$, over most of the Sun.
Inversion of this dependence
shows that in the convection zone rotation varies with
latitude approximately as on the surface,
with modest dependence on $r$.
Near the base of the convection zone there is a rapid transition,
over less than about $0.1 R$, to rotation depending little 
on $r$ and latitude in much of the radiative interior
({\eg} Thompson {\etal} 1996; Kosovichev 1996).

\begin{figure}[ht]
\epsfxsize=9cm\hspace*{1.8cm}\epsfbox{\figdir/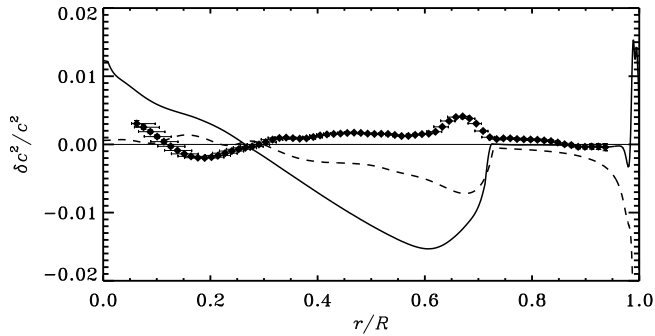}
\vskip -0.3cm
\caption{Relative differences in squared sound speed
between models with modified physics and the standard case,
in the sense (modified model) -- (standard model).
The solid line shows the effect of neglecting settling,
while the dashed line shows the effect
of using the Los Alamos Opacity Library rather than the OPAL tables.
Symbols show the inferred difference between the Sun and the
standard model, as in Fig.~1.}
\end{figure}

\section{Effects of modifying the physics}

To evaluate the significance of the comparatively close agreement
between the ``standard'' Model S and the Sun it is necessary to
consider other models with differing assumptions or physics.
%After all, similar agreement might be achievable with a broad range
%of differing physics, limiting the usefulness of the helioseismic
%results as constraints on stellar internal modelling.
%
An important example is the inclusion of settling and diffusion.
Figure~2 compares the difference between a model without
these effects, but using otherwise the same physics, and Model S,
with the difference between the Sun and Model S.
It is evident that neglect of settling would very considerably 
worsen the agreement between the Sun and the models
(see also Cox, Guzik \& Kidman 1989;
Christensen-Dalsgaard, Proffitt \& Thompson 1993;
Bahcall {\etal} 1997).
The figure also shows the effect of replacing the OPAL opacities
with the older Los Alamos Opacity Library (Huebner {\etal} 1977).
Clearly the revision of the opacity has improved the
agreement between the model and the Sun considerably,
although less so than
%It is interesting, however, that 
the inclusion of settling.
%has a somewhat larger effect than the change in the opacity.
This also suggests that the apparent improvement brought
about by including settling is not compromised by opacity
errors, as suggested by Elliott~(1995).

Very considerable effort has gone into work on the equation of state
({\eg} D\"appen 1992),
with corresponding improvements in the agreement between
the resulting models and the observed frequencies
%Even straight frequency comparisons have shown that
%equations of state representative of simple formulations
%often used in general stellar modelling
%is very far from being consistent with the helioseismic results
({\eg} Christensen-Dalsgaard, D\"appen \& Lebreton 1988).
Detailed analyses have
demonstrated the ability of the helioseismic data to
probe subtle properties of the thermodynamic state of the
solar plasma ({\eg} Christensen-Dalsgaard \& D\"appen 1992; 
Vorontsov, Baturin \& Pamyatnykh 1992; Elliott 1996);
this allows the use
of the solar convection zone, where the structure depends largely
on the equation of state, as a laboratory for plasma physics.

\section{What is wrong with the solar model?}

Despite the improvements in solar modelling 
and in the agreement between the model and the Sun,
the remaining highly significant discrepancy between the model and the Sun
indicates a lack in our understanding of stellar evolution;
although modest in the solar case, 
the error could quite possibly have more substantial effects in other stars
where conditions are more extreme than in the Sun.
%Thus there is clearly a case for concern.
%It is encouraging, however, that the data in this
%way promise insight beyond the simplicity of normal solar models.

\begin{figure}[ht]
\epsfxsize=12.8cm\hspace*{-0.3cm}\epsfbox{\figdir/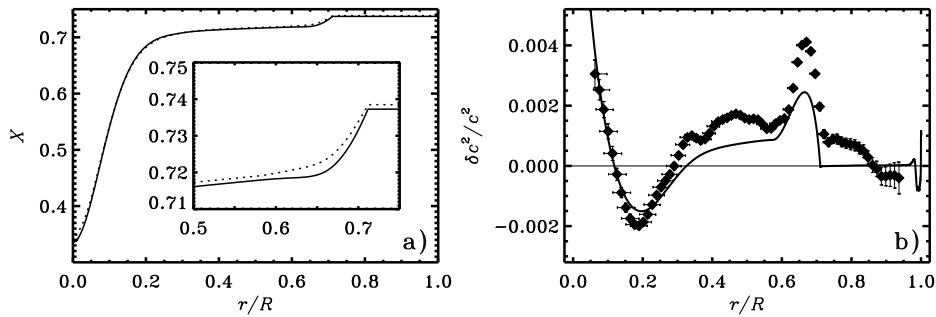}
\vskip -0.3cm
\caption{(a) Profiles of the abundance $X$ by mass of hydrogen.
The solid line shows the profile in
Model~S of Christensen-Dalsgaard {\etal} (1996),
whereas the dotted line shows a modified profile
aimed at trying to
match the sound-speed difference shown in Fig.~1
between the Sun and the model.
%The insert provides a blow-up of the region around the base of
%the convection zone.
(b) Difference in squared sound speed between the model
with modified $X$-profile and Model S.
The symbols show the inferred difference between the Sun
and Model S, as in Fig.~1.
Adapted from Bruntt (1996).
}
\end{figure}

The dominant features in the sound-speed difference
% , just beneath the convection zone and in the core,
occur in regions of the model with strong composition gradients ({\cf} Fig.~3),
resulting from nuclear burning in the core 
or helium settling from the convection zone.
These gradients would be affected by ``non-standard'' processes
causing mixing in convectively stable regions.
Mixing in the core could increase the hydrogen abundance at the centre
of the model while reducing it at the edge of the core;
the central sound speed would
therefore be increased, and the sound speed at the edge of the core
reduced, as required by Fig.~1.
Similarly, weak mixing beneath the convection zone would increase
the hydrogen abundance and sound speed in this region,
again potentially according for the observed bump.
As a toy model of such processes, Fig.~3 also shows 
an artificially modified hydrogen profile
% constrained to have approximately the same total 
%amount of hydrogen as the standard model.
and the corresponding change in the sound speed,
confirming 
%The sound-speed difference between this model and Model S,
%shown in Fig.~4, confirms
that redistribution of hydrogen can
in fact largely account for the observed behaviour.

Such suggestions evidently require physical mechanisms for the mixing.
Just beneath the convection zone 
the steep gradient in the helioseismically inferred rotation rate
is likely to be associated with circulation
which could cause mixing (Spiegel \& Zahn 1992; 
Gough {\etal} 1996; Elliott 1997).
Mixing might also be caused by instabilities associated
with the spin-down of the Sun from the usually assumed initial
state of rapid rotation ({\eg} Chaboyer {\etal} 1995),
or by penetration of convection beyond the unstable region.
%possibly in the form of intermittent deeply penetrating plumes.
Independent evidence for mixing beneath the convection zone
is provided by the destruction of lithium and beryllium
%as evidenced in the reduction of their solar photospheric
%abundances compared with the meteoritic values
({\eg} Chaboyer {\etal} 1995).

There appears to be no similarly simple explanation of potential
core mixing.
However, the Sun has
been shown to be unstable to low-order, low-degree g modes,
at least during earlier phases of its evolution
({\eg} Christensen-Dalsgaard, Dilke \& Gough 1974);
it is conceivable that the nonlinear development of these modes
can lead to mixing (Dilke \& Gough 1972).
The rotational spin-down 
might also lead to mixing of the core.
If such processes were to be common to low-mass
stars, they would have a substantial influence on our
understanding of stellar evolution, including an increase in the
estimated ages of globular clusters and hence in
the discrepancy with the cosmologically inferred age of the Universe.

Unfortunately, 
%such explanations in terms of changes in the
%composition profile are not unique.
substantial contributions to the difference between the
model and the Sun might come from perhaps less interesting
errors in the basic physics.
Indeed Tripathy {\etal} (1997) showed that the sound-speed
difference in Fig.~1 can be largely reproduced by modifications
to the opacity of less than about 5~\%.
While this is certainly smaller than the generally assumed uncertainty
in current opacity calculations, it remains to be seen whether
the specific change required is physically plausible.

Finally, I note that there is evidence for errors in the equation
of state in and below the helium ionization zone
({\eg} Dziembowski, Pamyatnykh \& Sienkiewicz 1992),
even when using the OPAL equation of state
%Indeed, it is possible to set up an inversion procedure
%such as to isolate the intrinsic error in $\Gamma_1$
%from errors in the model structure 
({\eg} Basu \& Christensen-Dalsgaard 1997).
The effect is small in the Sun, but it might be substantial 
in lower-mass stars where non-ideal plasma effects 
could be stronger.

\section{Relation to stellar astrophysics}

The helioseismic results clearly
give some confidence in modelling of stellar evolution.
%despite remaining problems discussed in the preceding section.
%Given the degree of detail obtained for the Sun one might 
%perhaps be tempted to wonder about the need for verification of
%stellar evolution theory as applied to other stars.
%This would be obviously foolish.
However, 
in part this relative success of the solar models undoubtedly stems from the
fact that the Sun is a comparatively simple type of star:
%even given the complexities of surface convection:
%at lower mass non-ideal effects in thermodynamics become
%substantial and molecular opacity is important;
for example, at only slightly higher mass than solar the problems of
a convective core would play a major role.
The ability to cover a broad range of parameters
makes investigations of other stars, ``classical'' as well as seismological,
an essential complement to the solar studies,
even though they can never be as detailed and precise as those obtained
for the Sun.
For example, discrepant period ratios in models of double-mode Cepheids
and $\delta$~Scuti stars
led to the prediction of a substantial increase in opacities
({\eg} Simon 1982; Andreasen \& Petersen 1988),
at temperatures in the range $10^5 - 10^6 \K$;
this falls within the solar convection zone
%such an opacity increase would have 
and hence would have no effect on solar structure.
The opacity increase in the OPAL
tables has in fact largely resolved this discrepancy
({\eg} Moskalik, Buchler \& Marom 1992).
%although problems may remain for the
%bump Cepheids (\note [ref to these proceedings??]).
Similarly, properties of convective cores,
including the important but highly uncertain
question of mixing beyond the region of instability,
might well be studied from observations of solar-like oscillations
in other stars (Kjeldsen, these proceedings)
or observation of sufficiently detailed spectra of oscillations
in, for example, $\delta$~Scuti or $\beta$~Cephei stars.
%Finally, observations of white dwarfs at different
%are providing quite detailed
%tests of the final phases of the evolution of moderate-mass stars
%({\eg} Kawaler 1995).

Independent stellar information may also help to compensate for
the non-uniqueness in the physical interpretation of the helioseismic data,
illustrated in the preceding section.
For example, investigations of element abundances
may provide further insight into the physics of 
mixing beneath stellar convection zones
({\eg} Baglin \& Lebreton 1990).

\section{Concluding remarks}

Major advances in helioseismology will result from
the extensive new data from the GONG network
({\eg} Harvey {\etal} 1996) and from the instruments on the SOHO satellite
({\eg} Scherrer {\etal} 1996),
as well as from the continued observations from other
ground-based instruments.
As a result, we shall be able to investigate 
in more detail solar structure, 
not least in the core and the convection zone,
as well as rotation and other aspects of solar internal dynamics.
In parallel with this, major advances in the
study of stellar oscillations may
%not least based on results
%from the recently approved French COROT mission
%({\eg} \note [ref]), 
lead to the definite detection 
of solar-like oscillations in other stars and a detailed
analysis of oscillations in $\delta$ Scuti stars and
other ``classical'' pulsators.
Finally, the new very large telescopes and
advances in stellar-atmosphere modelling (Gustafsson, these proceedings)
are likely to lead to major improvements in our knowledge about
stellar composition and the processes that control it, for
a variety of stars.

In this way we shall obtain much firmer tests of
stellar modelling, reaching beyond the 
fundamental properties of stars to the even more fundamental
questions of their physical basis.

\section*{Acknowledgements}

\footnotesize

I thank the authors of Basu {\etal} (1997) for permission
to quote results in advance of publication.
This work was supported in part by
the Danish National Research Foundation through its establishment of
the Theoretical Astrophysics Center.

{}

\end{document}